\documentclass[aip,jcp,amsmath,reprint]{revtex4-1}
\usepackage{graphicx}

\draft % marks overfull lines with a black rule on the right

\begin{document}

%\preprint{}

\title{A solvent-flux theory for nonequilibrium swelling dynamics of thermoresponsive microgels} %Title of paper



\author{Arturo Moncho-Jordá}
\email{moncho@ugr.es}
\thanks{Corresponding author}
\affiliation{Department of Applied Physics, University of Granada, 18071 Granada, Spain}
\affiliation{Institute
Carlos I of Theoretical and Computational Physics, University of Granada, 18071 Granada, Spain}

\author{Alessandro Patti}
\affiliation{Department of Applied Physics, University of Granada, 18071 Granada, Spain}
\affiliation{Institute
Carlos I of Theoretical and Computational Physics, University of Granada, 18071 Granada, Spain}

\author{Fabián A. García-Daza}
\affiliation{Faculty of Engineering, University of Deusto, Avda. Universidades, 24, Bilbao 48007, Spain}

\author{Alejandro Cuetos}
\email{acuemen@upo.es}
\thanks{Corresponding author}
\affiliation{Center for Nanoscience and Sustainable Technologies (CNATS), and Department of Physical, Chemical and Natural Systems, Universidad Pablo de Olavide, Sevilla, Spain}
%\author{}
%\thanks{}
%\affiliation{}


\date{\today}

\begin{abstract}
Thermoresponsive microgels undergo large reversible size changes as temperature modifies solvent quality. Predicting the nonequilibrium swelling and deswelling kinetics remains challenging because the polymer volume fraction, mechanical response, and solvent transport properties evolve during large volume changes. Here we develop a solvent-flux theory for the nonequilibrium dynamics of a spherical microgel. The radius-change rate is driven by the osmotic-pressure imbalance across the particle boundary and resisted by water transport through the polymer network, leading to a nonlinear equation for the global swelling coordinate and an explicit state- and temperature-dependent swelling diffusion coefficient, $D_{\mathrm{SW}}(\phi,T)$. In the linear-response regime, the theory recovers the Tanaka--Fillmore exponential relaxation, including relaxation-time scaling with the equilibrium radius and bulk modulus, $\tau_\mathrm{SW} \sim R_{\mathrm{eq}}^2\gamma/K_{\mathrm{eq}}$, while providing a microscopic interpretation of the polymer--solvent friction coefficient, $\gamma$. Beyond this limit, the model preserves quadratic size scaling while accounting for state-dependent transport and mechanical properties of the microgel. When applied to pNIPAM microgels, the model predicts asymmetric swelling and deswelling pathways, with deswelling occurring faster than swelling over the same temperature interval. Finite thermalization of the medium introduces a crossover from an intrinsic microgel-controlled regime to a thermalization-controlled regime, where the apparent relaxation time increases linearly with the external thermalization time and dynamic hysteresis-like loops appear in the radius--temperature plane. Finally, a stochastic extension based on the Smoluchowski equation predicts a transient broadening of the radius distribution during collapse, with maximum fluctuations in the volume-transition region. The theory therefore connects solvent transport, nonlinear swelling dynamics, finite thermalization effects, and nonequilibrium size fluctuations in responsive microgels.
\end{abstract}


\maketitle %\maketitle must follow title, authors, abstract and \pacs


\section{Introduction}

Stimuli-responsive microgels are cross-linked polymer networks whose size, internal polymer concentration, and mechanical properties can be reversibly modified by external stimuli such as temperature, pH, ionic strength, or solvent quality.~\cite{Murray1995,Saunders2009,Stuart2010} This coupling between molecular interactions, solvent transport, and network deformation underlies their use as model soft colloids and in applications involving
sensing, controlled release, separation, and catalysis.~\cite{Karg2019,Agrawal2018,Klinger2012,Kumari2023,Oberdisse2020,Welsch2011,Roa2017,Sabadasch2020} Among them, poly($N$-isopropylacrylamide) (pNIPAM) microgels are one of the most widely studied thermoresponsive systems because they undergo a reversible volume phase transition close to room temperature.~\cite{Heskins1968,Pelton2000,Halperin2018,Wu2003} Below the lower critical solution temperature (LCST), the polymer network is hydrated and swollen, whereas above the LCST the decrease in polymer--water affinity promotes solvent expulsion and particle collapse. This temperature-induced swollen-to-collapsed transition makes pNIPAM microgels a paradigmatic model system for studying the interplay between solvent transport, network elasticity, and nonequilibrium swelling/deswelling dynamics.

Although the equilibrium swelling of microgels can often be described using Flory--Rehner-type free-energy models,~\cite{Sierra-Martin2011b,Tanaka1978} predicting the kinetics by which a microgel evolves between two equilibrium states remains a nontrivial problem.~\cite{Tomari1994,Tomari1995,wahrmund2009,Bertrand2016,Nikolov2018,Camerin2018,Moreno2018,Song2019,Butler2022} Recent time-resolved experiments on pNIPAM-based nanogels have shown that collapse and swelling can occur on very different time scales, emphasizing the need for nonequilibrium descriptions that go beyond equilibrium swelling curves~\cite{Dallari2024}. This kinetics is relevant not only from a fundamental point of view, but also for applications in which the response time controls performance. In controlled release, for instance, temperature-induced changes in microgel volume modify the polymer volume fraction, mesh size, permeability, and partitioning of encapsulated molecules, thereby regulating the uptake or release of drugs, proteins, nanoparticles, or other active compounds.~\cite{Nolan2004,Serpe2005,Klinger2012,Trongsatitkul2013,Fundueanu2013,Vikulina2020} Similar considerations apply to microgel-based sensors, membranes, separation platforms, and microfluidic valves or actuators, where switching speed, reversibility, and dynamic hysteresis are determined by the swelling and deswelling kinetics.~\cite{Agrawal2018,DEramo2018} A predictive description of nonequilibrium swelling dynamics is therefore essential for connecting microscopic solvent transport through the polymer network with the macroscopic response of microgel-based functional materials.

The classical description of gel swelling kinetics was introduced by Tanaka and Fillmore.~\cite{Tanaka1979} In this framework, swelling is controlled by a collective diffusion process that couples network deformation and solvent permeation. For a spherical gel particle, the long-time relaxation of the radius is approximately single-exponential,
\begin{equation}
\label{eq:RtTanaka2}
\frac{R(t)-R_\mathrm{eq}}{R(t=0)-R_\mathrm{eq}}
\approx \frac{6}{\pi^2} \exp\left(-t/\tau_\mathrm{Tan}\right),
\end{equation}
where $R_\mathrm{eq}$ is the final equilibrium radius and $\tau_\mathrm{Tan}$ is the characteristic relaxation time, given by
\begin{equation}
\tau_\mathrm{Tan} =\frac{R_\mathrm{eq}^{2}}{\pi^2D_{\mathrm{coll}}}=\frac{R_\mathrm{eq}^{2}\gamma}{\pi^2K} .
\end{equation}
Here, $D_{\mathrm{coll}}\approx K/\gamma$ is the collective diffusion coefficient, where $K$ is the osmotic bulk modulus and $\gamma$ is the polymer--solvent friction coefficient per unit volume.~\cite{Tanaka1979} The resulting quadratic dependence of the relaxation time on particle size explains why microgels can respond much faster than macroscopic hydrogels and provides a standard route to extract $D_{\mathrm{coll}}$ from radius-versus-time measurements.~\cite{Suarez2006}

Despite its physical transparency, the Tanaka--Fillmore model is fundamentally a linear-response theory. It assumes small deformations around equilibrium, a homogeneous and isotropic network, and material parameters that remain constant during the relaxation. These assumptions become restrictive for large swelling or deswelling amplitudes, where the polymer volume fraction, osmotic bulk modulus, permeability, and polymer--solvent friction may all vary along the nonequilibrium pathway. In such cases, the effective mobility of the swelling coordinate is expected to be state dependent, giving rise to nonlinear and non-exponential relaxation.

In this work, we develop a nonlinear solvent-flux theory for the time-dependent radius of a spherical microgel. Instead of assuming a constant collective diffusion coefficient, we describe swelling and deswelling in terms of the net solvent flux through the microgel boundary, driven by the thermodynamic imbalance between the instantaneous microgel state and the surrounding solvent. The osmotic contribution to this driving force is obtained from the Flory--Rehner free energy, while the resistance to solvent transport is described through a state-dependent effective friction. Our aim is to derive an effective dynamical equation for the global swelling coordinate, $R(t)$, in terms of the free-energy landscape of the microgel. As we will show later, in the absence of thermal fluctuations, this dynamics can be written in the form
\begin{equation}
\label{eq:RtDSW}
\frac{dR}{dt} =-D_{\mathrm{SW}}(R,T)
\frac{\partial\beta F(R,T)}{\partial R},
\end{equation}
where $\beta=1/(k_{\mathrm B}T)$ and $F(R,T)$ is the instantaneous microgel free energy. The quantity $-\partial F/\partial R$ is the thermodynamic force conjugate to the swelling coordinate $R$: it drives the microgel downhill along the free-energy landscape towards the equilibrium radius. The coefficient $D_{\mathrm{SW}}(R,T)$ is the corresponding mobility, or effective diffusion coefficient, of the global swelling coordinate. It quantifies how efficiently this thermodynamic force is converted into changes of the particle radius. This coefficient should not be identified with either the molecular self-diffusion coefficient of water or the collective diffusion coefficient
$D_{\mathrm{coll}}$ of the Tanaka--Fillmore model, although the theory establishes explicit connections between these quantities.

%In the absence of thermal fluctuations, the radius obeys
%\begin{equation}
%\label{eq:RtDSW}
%\frac{dR}{dt}=-D_\mathrm{SW}(R,T)\frac{\partial\beta F(R,T)}{\partial R},
%\end{equation}
%where $\beta=1/(k_BT)$ and $F(R,T)$ is the instantaneous free energy. The coefficient $D_\mathrm{SW}(R,T)$ is an effective diffusion coefficient of the global swelling coordinate: it quantifies the mobility of the particle size along the free-energy landscape and should not be identified with either the molecular self-diffusion coefficient of water or the collective diffusion coefficient $D_{\mathrm{coll}}$. Although, as will be shown later, these coefficients are directly related.

We derive an explicit analytical expression for $D_{\mathrm{SW}}$ and apply the theory to the temperature-induced swollen-to-collapsed transition of pNIPAM microgels, as well as to the reverse collapsed-to-swollen process. In the linear-response regime, the model reduces to an exponential relaxation with the same quadratic size scaling as the Tanaka--Fillmore theory. Far from equilibrium, however, it accounts for the fact that both the mechanical response and the transport resistance evolve along the swelling or deswelling pathway.

Finally, we extend the deterministic description to include thermal fluctuations of the global swelling coordinate. This stochastic formulation, written in terms of the probability density $P(R,t)$, provides access not only to the mean radius but also to the variance and full time-dependent particle-size distribution after a temperature jump. This is particularly relevant near the volume-transition region, where the microgel becomes mechanically more susceptible to size fluctuations.

The remainder of the article is organized as follows. In Sec.~\ref{sec:theory}, we derive the solvent-flux model, obtain $D_{\mathrm{SW}}$, and discuss the linear-response limit. In Sec.~\ref{sec:experiments}, we introduce the pNIPAM microgel system and determine the thermodynamic and transport parameters entering the model. In Sec.~\ref{sec:results_T}, we analyze the predicted nonequilibrium swelling and deswelling dynamics under different temperature protocols. In Sec.~\ref{sec:theory_Smoluchowski}, we extend the theory using the Smoluchowski equation to describe the nonequilibrium size fluctuations. Finally, in Sec.~\ref{sec:conclusions}, we summarize the main conclusions.

\section{Theory}
\label{sec:theory}
\subsection{Solvent flow and microgel swelling}

We consider a single pNIPAM microgel particle dispersed in water at temperature $T$, with radius $R$ and internal polymer volume fraction $\phi$. Conservation of the total amount of polymer inside the particle implies that
\begin{equation}
\phi(R)=\phi_0\left(\frac{R_0}{R}\right)^3
\label{eq:phi}
\end{equation}
where $R_0$ is the microgel radius in the collapsed state and $\phi_0$ is the corresponding polymer volume fraction.

\begin{figure}[ht!]
	\centering
\includegraphics[width=1.0\linewidth]{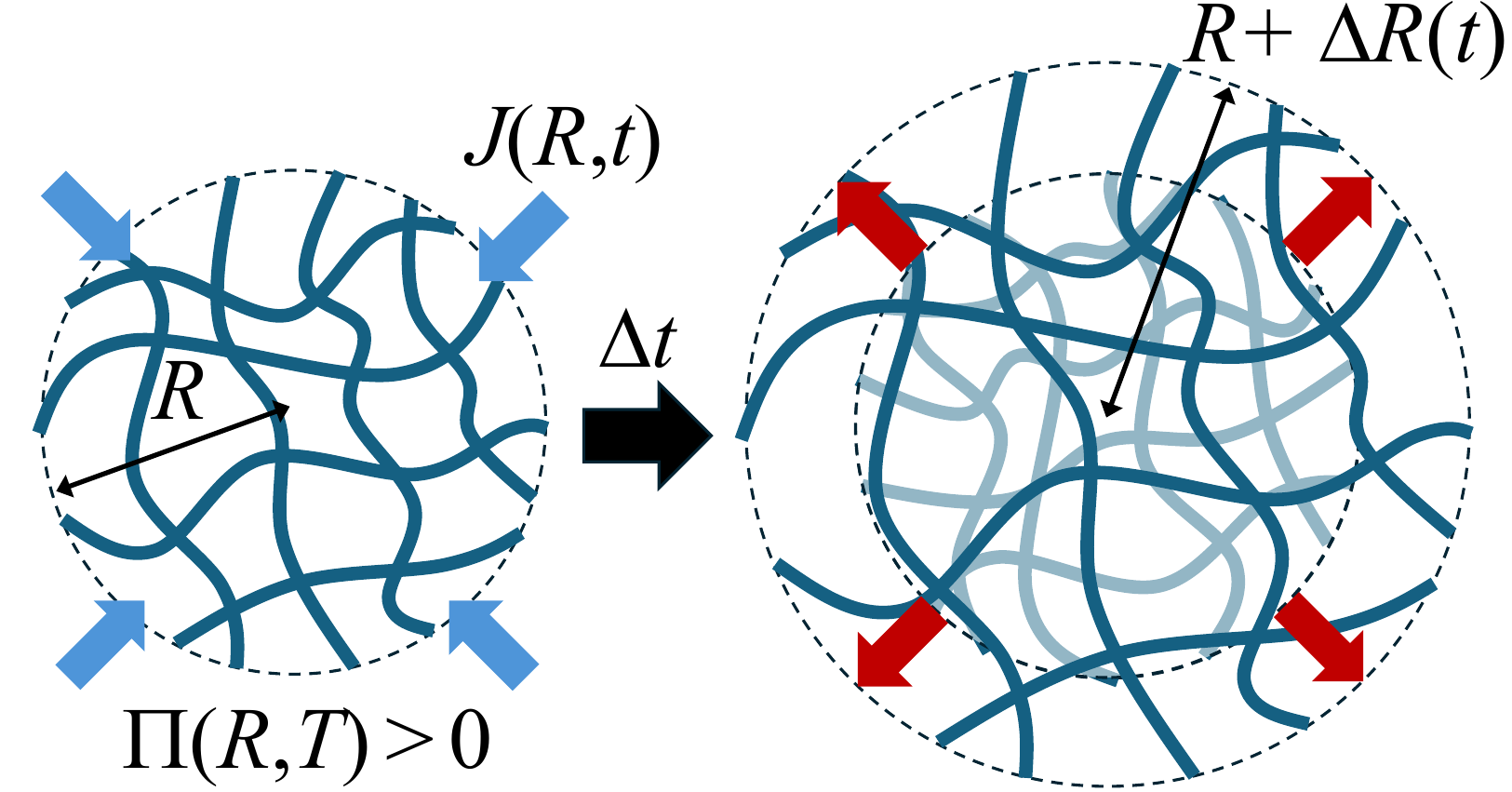}
	\caption{Schematic illustration of the swelling of a pNIPAM microgel driven by solvent uptake. The inward solvent flux is induced by a positive osmotic-pressure imbalance, $\Pi>0$, between the microgel interior and the surrounding bulk solution,leading to an increase of the microgel radius. The opposite case, $\Pi<0$, corresponds to deswelling, where solvent is expelled from the microgel and the particle radius decreases.}
    \label{fig:scheme_swelling}
\end{figure}

We now consider the change in microgel size induced by a variation of the temperature of the surrounding medium from $T$ to $T+\Delta T$. For definiteness, we first focus on the case $\Delta T<0$. Under these conditions, the solvent quality improves and the pNIPAM microgel swells. This swelling process is associated with an inward flux of water molecules across the particle surface at $r=R$. Assuming that solvent transport through the polymer network is overdamped and diffusion-controlled, the local water flux can be written in the standard Smoluchowski drift--diffusion form:~\cite{Dhont1996}
\begin{equation}
\mathbf{J}(\mathbf{r},t)=-D_\mathrm{w} \big( \nabla c(\mathbf{r},t) - c(\mathbf{r},t)\beta\mathbf{f}(\mathbf{r},t)\big).
\label{eq:J}
\end{equation}
where $\beta=1/(k_{\mathrm B}T)$, $D_{\mathrm w}$ is the diffusion coefficient of water inside the microgel, and $c(\mathbf{r},t)$ is the local concentration of solvent molecules at position $\mathbf{r}$ and time $t$. The first term in the right-hand side of Eq.~\eqref{eq:J} accounts for ordinary diffusion, whereas the second term describes the drift of solvent molecules induced by the force acting on a single water molecule, $\mathbf{f}(\mathbf{r},t)$. In the present case, this force represents the effective interaction exerted by the polymer network.

Eq.~\eqref{eq:J} can be simplified by noting that water behaves as an effectively incompressible solvent, so that its number density remains essentially constant during the swelling process, i.e., $c(\mathbf{r},t)=c_\mathrm{w}=33.3$~molec$/$nm$^3$. This approximation corresponds to a quasi-static regime in which local density fluctuations relax much faster than the characteristic swelling time. Under these conditions, the dynamics is governed by solvent transport across the microgel interface rather than by density variations within the solvent bulk. We take the center of the microgel as the origin, $r=0$, so that the particle boundary is a spherical surface of radius $R$. The total radial solvent flux across this surface is then given by
\begin{equation}
J(R,t)=D_\mathrm{w}c_\mathrm{w}\beta f(R,t).
\label{eq:J3}
\end{equation}
where $J(R,t)$ and $f(R,t)$ are, respectively, the radial flux and radial external force at the microgel surface $r=R$.

The rate of change of the number of solvent molecules inside the microgel is obtained from the flux as
\begin{equation}
\frac{dN_\mathrm{w}}{dSdt}=-J(R,t) \ \ \rightarrow \ \ \frac{dN_\mathrm{w}}{dt}=-4\pi R^2J(R,t).
\label{eq:dNdt}
\end{equation}

For a spatially homogeneous microgel, the total number of solvent molecules inside the particle is obtained from Eq.~\eqref{eq:phi} as
\begin{equation}
    N_\mathrm{w}=c_\mathrm{w}(1-\phi)V=\frac{4}{3}\pi c_\mathrm{w}( R^3-\phi_0R_0^3).
    \label{eq:NwR}
\end{equation}
where $V=(4/3)\pi R^3$ is the total volume of the particle and the factor $1-\phi$ accounts for the volume fraction available to the solvent inside the microgel. Since $c_{\mathrm w}$ is constant, changes in the number of solvent molecules inside the microgel arise solely from changes in the microgel radius:
\begin{equation}
\frac{dN_\mathrm{w}}{dt}=c_\mathrm{w}4\pi R^2\frac{dR}{dt}.
\label{eq:dNdt2}
\end{equation}

Combining Eq.~\eqref{eq:dNdt} and Eq.~\eqref{eq:dNdt2} and then using Eq.~\eqref{eq:J3}, we find that
\begin{equation}
\frac{dR}{dt}=-\frac{J(R,t)}{c_\mathrm{w}}=-D_\mathrm{w}\beta f(R,t)
\label{eq:dRdt}
\end{equation}

If the average force acting on the molecules is attractive, this leads to an increase of the radius (swelling). Conversely, for a repulsive force, the radius decreases (deswelling).

The next step is to relate the average radial force acting on a water molecule at the microgel interface, $f(R,t)$, to the osmotic-pressure difference, $\Pi=\Pi_{\mathrm{in}}-\Pi_{\mathrm{out}}$, where $\Pi_\mathrm{in}$ and $\Pi_\mathrm{out}$ are the pressures inside and outside the microgel. The osmotic pressure $\Pi$ represents a mechanical force per unit area exerted by the polymer network on the solvent. However, in order to determine the force acting on a single water molecule crossing the microgel surface, one must relate the osmotic pressure to the corresponding change in chemical potential. For an incompressible solvent, the change in chemical potential is related to the pressure variation through
\begin{equation}
d\mu_\mathrm{w} = v_\mathrm{w}\, d\Pi,
\end{equation}
where $v_\mathrm{w}$ is the molecular volume of water, given by $v_\mathrm{w} = 1/c_\mathrm{w}$. If the osmotic imbalance across the microgel interface is $\Pi$, the corresponding jump in solvent chemical potential per molecule is $\Delta \mu_\mathrm{w} \simeq (1/c_\mathrm{w})\Pi$. This quantity represents the reversible work per molecule associated with crossing the interface. The microscopic force acting on a solvent molecule is given by the gradient of the chemical potential,
\begin{equation}
\mathbf{f} = - \nabla \mu_\mathrm{w}.
\end{equation}

We assume that the chemical-potential variation takes place across a region of thickness $\xi$, whose magnitude is of the order of the network mesh size. For the cross-linked polymer network considered here, $\xi$ is identified with the average distance between neighboring crosslinks. The magnitude of the resulting effective force acting on a solvent molecule can then be estimated as
\begin{equation}
f\sim -\frac{\Delta \mu_\mathrm{w}}{\xi}=-\frac{\Pi}{c_\mathrm{w}\,\xi}.
\label{eq:fR}
\end{equation}
$\Pi>0$ corresponds to an outward mechanical stress exerted by the polymer network on the solvent, which drives solvent uptake and therefore swelling of the microgel. Conversely, $\Pi<0$ corresponds to deswelling conditions.

Eq.~\eqref{eq:fR} states that the force acting on a single solvent molecule crossing the microgel boundary is controlled by the osmotic pressure and inversely proportional to both the solvent number density and the effective interfacial thickness. Combining Eq.~\eqref{eq:dRdt} and \eqref{eq:fR}, we deduce that the swelling kinetics is therefore governed by the nonlinear ordinary differential equation
\begin{equation}
\frac{dR}{dt}=- D_\mathrm{w}\,\beta\, f(R)=
 \frac{D_\mathrm{w}}{c_\mathrm{w}\,\xi}\,\,\beta\Pi.
\label{eq:dRdt2}
\end{equation}

Finally, we consider that the swelling of the microgel is approximately affine. Under this assumption, the average distance between neighboring crosslinks scales linearly with the microgel radius and can be written as
\begin{equation}
    \xi(R) = \frac{\xi_\mathrm{ref}}{R_\mathrm{ref}}R = \lambda R,
\end{equation}
where $\xi_\mathrm{ref}$ and $R_\mathrm{ref}$ are, respectively, the crosslink-to-crosslink distance and the microgel radius in a reference state, and $\lambda \equiv\xi_\mathrm{ref}/R_\mathrm{ref}$ is a dimensionless constant parameter. We use the swollen state as the reference state, in which the polymer chains are relatively extended and the average distance between neighboring
crosslinks can be estimated as
\begin{equation}
\xi_{\mathrm{ref}}\simeq (n+1)l_0,
\end{equation}
where $l_0\approx 0.25$~nm is the effective length of a monomeric NIPAM unit and $n$ is the average number of monomers between consecutive crosslinks. Substituting $\xi(R)=\lambda R$ into the solvent-flux expression gives
\begin{equation}
\frac{dR}{dt}=\frac{D_\mathrm{w}(\phi,T)}{c_\mathrm{w}\,\lambda R}\,\,\beta\Pi(\phi,T).
\label{eq:dRdt3}
\end{equation}

In this last equation, we explicitly account for the fact that the osmotic pressure depends on both $\phi$ and $T$, and that the diffusion coefficient of the solvent inside the microgel is smaller than in bulk solution due to friction with the fluctuating polymer chains, so $D_\mathrm{w}=D_\mathrm{w}(\phi,T)$. Eq~\eqref{eq:dRdt3} provides a theoretical description of the temperature-driven swelling and deswelling dynamics of a single microgel particle. After a change in temperature, the osmotic-pressure imbalance becomes nonzero and drives solvent uptake or release. The microgel radius then evolves until the new equilibrium state is reached, characterized by $\Pi=0$.

To close the theory, we now require explicit analytical expressions for $\Pi(\phi,T)$ and $D_\mathrm{w}(\phi,T)$.

\subsection{Diffusion coefficient of water molecules inside the microgel}

For water molecules inside the microgel, the mobility is reduced with respect to bulk water by the presence of the polymer network. We describe this reduction directly through the state- and temperature-dependent water diffusion
coefficient,
\begin{equation}
D_{\mathrm w}(\phi,T) =
\frac{k_{\mathrm B}T}{6\pi\eta(T)a_{\mathrm w}}
\left( \frac{1-\phi}{1+\phi} \right)^2 \exp\left[
-b\frac{\phi}{1-\phi} \right],
\label{eq:Dw_eff}
\end{equation}
where $a_{\mathrm w}=0.1066~\mathrm{nm}$ is the hydrodynamic radius of a water molecule and $\eta(T)$ is the viscosity of bulk water. The latter is described by the empirical interpolation formula
\begin{equation}
\eta(T)=\left(2.414\times10^{-5}\right)10^{247.8/(T-140)},
\label{eq:water_viscosity}
\end{equation}
with $T$ expressed in Kelvin and $\eta$ in $\mathrm{Pa\,s}$.

Eq.~\eqref{eq:Dw_eff} separates the reduction of water mobility inside the microgel into two contributions. The prefactor $k_{\mathrm B}T/(6\pi\eta(T)a_{\mathrm w})$ is the Stokes--Einstein estimate for the self-diffusion coefficient of bulk water at temperature $T$. The algebraic factor $((1-\phi)/(1+\phi))^2$ is the Mackie--Meares obstruction factor and accounts for the reduction of available diffusion pathways as the polymer volume fraction increases.~\cite{Mackie1955,Kanduc2021} The exponential term $\exp(-b\phi/(1-\phi))$ introduces an additional phenomenological free-volume correction, inspired by free-volume models of diffusion.~\cite{Yasuda1968} This term captures the stronger decrease of water mobility in polymer-rich states, with the dimensionless parameter $b$ controlling the strength of this contribution.

In the dilute limit, $\phi\to 0$, both correction factors tend to unity and Eq.~\eqref{eq:Dw_eff} reduces to the bulk-water Stokes--Einstein diffusion coefficient.

The reduction of water mobility inside the microgel can be interpreted in terms of an effective viscosity, $\eta_{\mathrm{eff}}(\phi,T)$,
defined as
\begin{equation}
\eta_{\mathrm{eff}}(\phi,T)=\eta(T)\left(\frac{1+\phi}{1-\phi}
\right)^2\exp\left[b\frac{\phi}{1-\phi}
\right].
\label{eq:eta_eff}
\end{equation}

\subsection{Free energy, osmotic pressure and bulk modulus}

The equilibrium swelling of cross-linked polymer networks is commonly described within the Flory--Rehner framework, which combines the Flory--Huggins free energy of polymer--solvent mixing with the elastic free energy of an affine polymer network and, in charged microgels, an additional ionic contribution.~\cite{FloryRehner1943a,FloryRehner1943b,Sierra-Martin2011b,Tanaka1980} This approach has been widely used to describe the volume phase transition and swelling curves of pNIPAM-based microgels.~\cite{Fernandez-Barbero2002,Sierra-Martin2011,Moncho-Jorda2016,Lopez2017,Nigro2017,Sbeih2019,moncho-jorda2025} Polymer--solvent mixing favors solvent uptake under good-solvent conditions, whereas the cross-linked network opposes large deformations through its elastic free energy. Additional osmotic pressure from confined counterions can further promote swelling in charged systems. The explicit expression of the Flory--Rehner microgel free energy is given by:
\begin{equation}
\label{freeenergyfinal2}
\begin{aligned}
F(\phi,T)&=N_{\mathrm{ch}}k_\mathrm{B}T\Bigg[\frac{3}{2}\left(
\left(\frac{\phi_{0}}{\phi}\right)^{2/3}
-\frac{1}{3}\ln\!\left(\frac{\phi_{0}}{\phi}\right)
-1\right) \\
&\quad
+\, \ln \phi + nB \left(\frac{1}{\phi}-1\right)\ln(1-\phi)  \\
&\quad
+\, nB \chi(T) (1-\phi) + n_\mathrm{e} \ln\!\left(\frac{\phi}{\phi_{0}}\right)
\Bigg].
\end{aligned}
\end{equation}
where $N_\mathrm{ch}$ is the total number of chains in the microgel, $\phi_0$ is the polymer volume fraction in the collapsed state, $n$ is the average number of monomers per chain, $n_\mathrm{e}$ is the average number of charged monomers per chain, $B=v_\mathrm{mon}/v_\mathrm{w}=3.644$ is the ratio between the monomer volume and the solvent volume,~\cite{moncho-jorda2025} and $\chi(T)$ is the Flory--Huggins solvency parameter. For pNIPAM microgels, $\chi(T)$ is of the order of $0$--$0.2$ in the swollen, low-temperature regime and increases with temperature, reflecting the progressive worsening of solvent quality above the LCST.

The osmotic pressure is given by $\Pi=-c_{\mathrm w}(\partial F/\partial N_{\mathrm w})_T$. Using Eq.~\eqref{eq:NwR} to express this derivative in terms of $R$, we obtain
\begin{equation}
\Pi =-\frac{1}{4\pi R^2}\frac{d F}{dR},
\label{eq:PiF}
\end{equation}
which leads to
\begin{align}
\Pi(\phi,T)=&\frac{N_\mathrm{ch}k_\mathrm{B}T}{V_0}\bigg[ \Big(\frac{3}{2}+n_\mathrm{e}-nB\Big)\frac{\phi}{\phi_0}-\Big(\frac{\phi}{\phi_0}\Big)^{1/3} \nonumber \\
-&\frac{nB}{\phi_0}\Big( \ln(1-\phi) +\chi(T)\phi^2\Big)\bigg],
\label{eq:PiFloryRehner}
\end{align}
where $V_0=(4/3)\pi R_0^3$ is the volume of the microgel in the collapsed state. The final equilibrium state is achieved when $\Pi=0$, which corresponds to a minimum in the intrinsic free energy, $F(R,T)$.

The bulk modulus of the microgel, which provides the mechanical rigidity of the polymer network, is obtained as $K=\phi (\partial \Pi /\partial \phi)_T$, leading to
\begin{align}
K(\phi,T)=&\frac{N_\mathrm{ch}k_\mathrm{B}T}{V_0}\bigg[ \Big(\frac{3}{2}+n_\mathrm{e}-nB\Big)\frac{\phi}{\phi_0}-\frac{1}{3}\Big(\frac{\phi}{\phi_0}\Big)^{1/3} \nonumber \\
+&\frac{nB}{\phi_0}\Big( \frac{\phi}{1-\phi} -2\chi(T)\phi^2\Big)\bigg].
\label{eq:KFloryRehner}
\end{align}

\subsection{Calculation of $D_\mathrm{SW}$}

Once the mechanical properties of the microgel have been described within the Flory--Rehner framework, we now connect the solvent-flux formulation to the coarse-grained dynamics of the microgel radius, Eq.~\eqref{eq:RtDSW}, by rewriting the osmotic pressure in terms of the derivative of the microgel free energy. Inserting Eq.~\eqref{eq:PiF} into Eq.~\eqref{eq:dRdt3}, together with $R^3=3\phi_0V_0/(4\pi\phi)$, gives
\begin{equation}
\frac{dR}{dt}= -\frac{D_\mathrm{w}(\phi,T)\phi}{3c_\mathrm{w}\lambda \phi_0 V_0}\frac{d\beta F(R,T)}{dR} .
\label{eq:dRdtfinal3}
\end{equation}

Comparing with Eq.~\eqref{eq:RtDSW}, we identify the state- and temperature-dependent swelling diffusion coefficient as
\begin{equation}
D_\mathrm{SW}(R,T)=\frac{D_\mathrm{w}(\phi,T)\phi}{3c_\mathrm{w}\lambda \phi_0 V_0},
\label{eq:DSW}
\end{equation}
so
\begin{equation}
\frac{dR}{dt}=M(R,T)\equiv-D_\mathrm{SW}(R,T)\frac{d\beta F(R,T)}{dR},
\label{eq:dRdtM}
\end{equation}
where $M(R,T)$ denotes the instantaneous swelling or deswelling rate.

\subsection{Comparison with the Tanaka--Fillmore model}

After deriving the effective swelling diffusion coefficient, $D_{\mathrm{SW}}$, it is important to examine the connection between the present flux-based dynamics and the classical linear description of gel swelling. In that framework, the characteristic relaxation time scales quadratically with particle size and is inversely proportional to the collective diffusion coefficient, namely $\tau_{\mathrm{Tan}} = R_{\mathrm{eq}}^2\gamma/(\pi^2K)$ .~\cite{Tanaka1979,matsuo1988,Suarez2006,wahrmund2009,Dallari2024} We now show that our model recovers this scaling in the linear-response regime, while extending it to situations in which the mechanical and transport properties evolve during the swelling or deswelling process. To obtain this limit, we consider small deviations from equilibrium. Under these conditions, the polymer volume fraction can be approximated by its equilibrium value, $\phi\approx\phi_{\mathrm{eq}}$, and the state-dependent transport coefficients can be evaluated at equilibrium. The osmotic pressure can then be linearized around the equilibrium radius. Thus, near $R=R_{\mathrm{eq}}$, one has
\begin{equation}
\Pi(R)\approx \left(\frac{\partial \Pi}{\partial R}\right)_{R_\mathrm{eq}}(R-R_\mathrm{eq}).
\end{equation}

We can write $\partial \Pi/\partial R$ in terms of the bulk modulus:
\begin{equation}
    \frac{\partial \Pi}{\partial R}=\frac{\partial \Pi}{\partial \phi}\frac{\partial \phi}{\partial R}=-\frac{3}{R}\phi\frac{\partial \Pi}{\partial \phi}=-\frac{3}{R}K,
\end{equation}
where we used the definition of the bulk modulus, namely $K=\phi(\partial \Pi/\partial \phi)_T$. Therefore
\begin{equation}
\Pi(R)\approx -3\frac{K_\mathrm{eq}}{R_\mathrm{eq}}(R-R_\mathrm{eq}),
\label{eq:Pieq}
\end{equation}
where $K_\mathrm{eq}$ is the bulk modulus in the equilibrium state.

Inserting Eq.~\eqref{eq:Pieq} into Eq.~\eqref{eq:dRdt3}, we find the following differential equation, valid only close to the equilibrium state:
\begin{equation}
\frac{dR}{dt}=-\frac{1}{\tau_\mathrm{SW}} (R- R_\mathrm{eq}),
\label{eq:dRdta}
\end{equation}
where $\tau_\mathrm{SW}$ is the corresponding characteristic swelling time, defined as
\begin{equation}
\tau_\mathrm{SW} \equiv \frac{2\pi \eta_\mathrm{eff}(\phi_\mathrm{eq},T)a_\mathrm{w}c_\mathrm{w}\lambda R_\mathrm{eq}^2}{K_\mathrm{eq}}.
\label{eq:tauSW}
\end{equation}

Integrating Eq.~\eqref{eq:dRdta} from the initial state at $t=0$, with microgel radius $R(t=0)$, to an arbitrary time $t$ at which the radius is $R(t)$ leads to
\begin{equation}
\frac{R(t)-R_\mathrm{eq}}{R(t=0)-R_\mathrm{eq}}=\exp(-t/\tau_\mathrm{SW}).
\end{equation}

As observed, our flux-based and the Tanaka--Fillmore models both lead to the same common result in conditions close to equilibrium. On the one hand, comparing the previous equation with the leading order predicted in the Tanaka-Fillmore model (see Eq.~\eqref{eq:RtTanaka2}), we find the same exponential behavior. On the other hand, the deduced expression for $\tau_\mathrm{SW}$ in Eq.~\eqref{eq:tauSW} shares the expected scaling, namely $\tau_\mathrm{SW} \sim R_\mathrm{eq}^2/K_\mathrm{eq}$. In addition, by identifying $\tau_\mathrm{SW}=\tau_\mathrm{Tan}=R_\mathrm{eq}^2\gamma/(\pi^2K_\mathrm{eq})$, we find an explicit expression for the friction coefficient per unit of volume appearing in the Tanaka-Fillmore model:
\begin{equation}
    \gamma  = 2\pi^3 \eta_\mathrm{eff}(\phi_\mathrm{eq},T)a_\mathrm{w}c_\mathrm{w}\lambda .
\end{equation}

This expression provides a microscopic interpretation of the phenomenological friction coefficient $\gamma$, which can be viewed as the collective dissipation associated with solvent transport through the polymer network during swelling or deswelling. In this picture, the solvent experiences an effective viscous resistance controlled by the local polymer environment, represented by the effective viscosity $\eta_{\mathrm{eff}}$. The factor $\lambda$, defined through $\xi(R)=\lambda R$, accounts for the relative thickness of the interfacial region over which the chemical-potential imbalance is converted into a force acting on the solvent molecules. A larger value of $\lambda$ corresponds to a larger distance over which the chemical-potential change is distributed, leading to a weaker driving force per solvent molecule and, consequently, to a slower swelling response.

Finally, we find the following expression for the collective diffusion coefficient defined in the framework of the Tanaka--Fillmore theory:
\begin{equation}
D_\mathrm{coll}=\frac{3\beta K_\mathrm{eq}D_\mathrm{w}(\phi_\mathrm{eq},T)}{\pi^2c_\mathrm{w}\lambda}=\frac{9\beta K_\mathrm{eq}\phi_0V_0}{\pi^2\phi_\mathrm{eq}}D_\mathrm{SW}(\phi_\mathrm{eq},T).
\label{eq:Dcollmodel}
\end{equation}

Consequently, our model captures the main trends and physical dependencies, while additionally accounting for the fact that the mechanical properties and friction coefficient of the microgel evolve during the swelling process.

\section{Experimental system}
\label{sec:experiments}

We apply the model to pNIPAM microgels synthesized in Milli-Q water by precipitation polymerization, following the protocol described by Rubio-Andrés et al.~\cite{Synthesis2025} The particles were prepared using N-isopropylacrylamide as monomer, N,N'-methylene-bisacrylamide (BIS) as crosslinker, and potassium persulfate as initiator, with a crosslinking density of $4.8~\mathrm{mol}\%$. From the synthesis conditions and the particle concentration obtained by nanoparticle tracking analysis, we estimate an average number of monomers per chain $n\simeq 9.949$. The initiator also introduces a small charge, corresponding to an average number of charged monomers per chain $n_\mathrm{e}=0.107$.

The hydrodynamic radii of the microgel in its swollen and collapsed states were experimentally determined \textit{via} Dynamic Light Scattering (Malvern Zetasizer NanoZ), yielding $R_\mathrm{swollen} = (345 \pm 25)~\mathrm{nm}$ at $T = 15^\circ$C and $R_0 = (200 \pm 10)~\mathrm{nm}$ at $T = 48^\circ$C. The corresponding volume of the microgel in the collapsed state is $V_0=3.35 \times 10^7$~nm$^3$. We take the polymer volume fraction in the collapsed state to be $\phi_0=0.5$, which is consistent with experimental estimates for collapsed pNIPAM microgels, typically lying in the range $\phi_0\simeq 0.4$--$0.6$ depending on particle architecture and crosslinker chemistry.~\cite{Sbeih2019,Deen2015} Using the radius in the swollen state, we estimate that $\lambda=(n+1)l_0/R_\mathrm{swollen}=0.008$.

\begin{figure}[ht!]
	\centering
\includegraphics[width=1.0\linewidth]{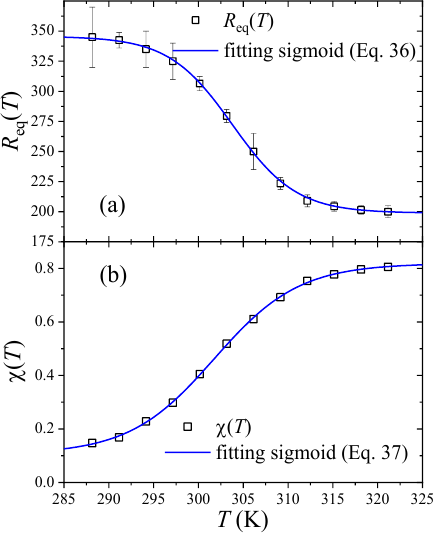}
	\caption{(a) Equilibrium hydrodynamic microgel radius as a function of $T$. (b) Flory--Huggins solvency parameter as a function of $T$. Blue lines show the corresponding fits of $R_\mathrm{eq}(T)$ and $\chi(T)$ using Eqs.~\eqref{eq:sigmoid1} and \eqref{eq:sigmoid2}.}
    \label{fig:R_chi}
\end{figure}

The full swelling behavior of the thermo-responsive pNIPAM microgels obtained through DLS is represented by square symbols in Fig.~\ref{fig:R_chi}(a), which plots the mean hydrodynamic radius as a function of temperature. As observed, the particle size exhibits the expected behavior: at low temperatures, the microgels are in a swollen state, with the size reaching a plateau well below the transition temperature. As the temperature increases, the size decreases, eventually stabilizing at a new plateau corresponding to the collapsed state. The experimental data are well described by the following sigmoid function, shown as a solid line in Fig.~\ref{fig:R_chi}(a):
\begin{equation}
	\label{eq:sigmoid1}
	R(T)=A_0+\frac{A_1-A_0}{1+\exp \big((T-T_\mathrm{LCST})/\Delta \big)},
\end{equation}
where $A_1=(345.72\pm 1.18)$~nm, $A_0=(198.89\pm 1.03)$~nm, $T_\mathrm{LCST}=(303.79\pm 0.12)$~K and $\Delta=(3.56 \pm 0.12)$~K. The resulting LCST transition temperature is close to the reported value for pNIPAM microgels.~\cite{Fernandez-Barbero2002,Lopez2017,Scotti2019}

The Flory--Huggins solvency parameter, $\chi$, was determined by reproducing the experimentally observed swelling behavior. For this purpose, we apply internal mechanical equilibrium condition $\Pi = 0$  in Eq.~\eqref{eq:PiFloryRehner} for each radius $R$ measured \textit{via} DLS. The resulting values of $\chi$ are plotted as square symbols in Fig.~\ref{fig:R_chi}(b). As expected for pNIPAM microgels, $\chi$ increases with temperature, reflecting the increasing hydrophobicity of the polymer chains above the LCST, which drives the deswelling process. The obtained values of $\chi$ lie within the range of commonly accepted values, confirming that the theoretical model provides a reliable description of the system. To analytically reproduce the extracted values of $\chi(T)$, we fit the data using an empirical sigmoid function (see the blue solid line in Fig.~\ref{fig:R_chi}(b)), given by:
\begin{equation}
	\chi (T)=\chi_1+\frac{\chi_0-\chi_1}{1+\exp(-\kappa(T-T^*))}.
    \label{eq:sigmoid2}
\end{equation}

The fit yields the following parameter values: $\chi_1 = 0.106 \pm 0.008$, $\chi_0 = 0.189 \pm 0.006$, $T^* = (301.76 \pm 0.14)$~K. $\kappa = (0.211 \pm 0.007)$~K$^{-1}$. As observed, the resulting crossover temperature $T^*$ is very close to $T_\mathrm{LCST}$.

The number of polymer chains that effectively contribute to the elastic response of the microgel is not directly measured in the present work. Instead, it is fixed by calibrating the model against a representative collapsed-state bulk modulus. Motivated by osmotic-compression measurements of individual pNIPAM microgels, which report bulk moduli in the $10^2~\mathrm{kPa}$ range in the collapsed regime,~\cite{Sierra-Martin2011} we impose $K_{\mathrm{eq}}=200~\mathrm{kPa}$ at the collapsed equilibrium state. This calibration gives $N_{\mathrm{ch}}\approx1.82\times10^5$. Using the same value of $N_{\mathrm{ch}}$, the model predicts $K_{\mathrm{eq}}\simeq14.7~\mathrm{kPa}$ in the swollen state.

Before discussing the nonequilibrium swelling dynamics, we specify the remaining parameter $b$ entering the effective-viscosity expression, Eq.~\eqref{eq:eta_eff}. In the dilute limit, $\phi\to0$, the model recovers
the bulk-water Stokes--Einstein diffusion coefficient. For example, at $T=298~\mathrm K$ we obtain $D_{\mathrm w}(\phi=0,T=298~\mathrm K)= 2.3\times10^{-9}~\mathrm{m^2\,s^{-1}}$, in agreement with experimental measurements of the self-diffusion coefficient of bulk water.~\cite{Mills1973} The parameter $b$ controls the additional reduction of water mobility in polymer-rich environments. We calibrate it using the molecular-dynamics simulations of Kanduc et al.,~\cite{Kanduc2021} who reported $D_{\mathrm w}(\phi=0.8,T=340~\mathrm K)=
3\times10^{-11}~\mathrm{m^2\,s^{-1}}$ for water inside collapsed pNIPAM. This condition gives $b\simeq0.208$. With this value, the model predicts
$D_{\mathrm w}(\phi=0.5,T=313~\mathrm K)= 2.97\times10^{-10}~\mathrm{m^2\,s^{-1}}$ for a collapsed pNIPAM microgel, which is about one order of magnitude smaller than the corresponding bulk value and is consistent with experimental observations of reduced water mobility in collapsed microgels.~\cite{Sierra-Martin2005} This reduction
reflects the hindered diffusion of water through the polymer-rich microgel interior.

As a final consistency check, we use Eq.~\eqref{eq:Dcollmodel} to estimate the collective diffusion coefficient. The resulting values lie in the range $D_{\mathrm{coll}}\simeq5\times10^{-12}$--$10^{-10}~ \mathrm{m^2\,s^{-1}}$, in agreement with reported experimental values for swelling and deswelling dynamics of pNIPAM gels and microgels.~\cite{Andersson1998,Tanaka1986,Suarez2006}

\section{Nonequilibrium swelling behavior under temperature protocols}
\label{sec:results_T}
\subsection{Nonequilibrium response to a sudden temperature jump}

We begin the results section by considering a simple temperature-jump protocol. Initially, the microgel is at equilibrium in the swollen state at $T_{\mathrm{ini}}=288~\mathrm{K}$, with an initial radius
$R(t=0)=345~\mathrm{nm}$. The temperature of the surrounding medium is then suddenly changed to $T_{\mathrm f}=321~\mathrm{K}$. This abrupt temperature increase drives the system out of equilibrium and induces the deswelling of the microgel until a new collapsed equilibrium state is reached, with $R_{\mathrm{eq}}=200~\mathrm{nm}$.

\begin{figure}[ht!]
	\centering
\includegraphics[width=1.0\linewidth]{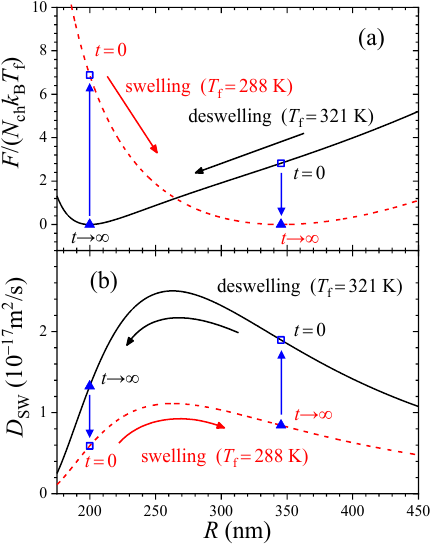}
	\caption{(a) Normalized free-energy landscape of the microgel as a function of the particle radius $R$. (b) State-dependent swelling diffusion coefficient, $D_{\mathrm{SW}}(R,T_\mathrm{f})$, as a function of $R$. In both panels, the black solid line corresponds to the deswelling process after a temperature jump to $T_{\mathrm f}=321~\mathrm{K}$, whereas the red dashed line corresponds to the swelling process after a temperature jump to $T_{\mathrm f}=288~\mathrm{K}$. Blue squares denote initial state, whereas blue triangles show the final equilibrium state.}
    \label{fig:F_DSW}
\end{figure}

Fig.~\ref{fig:F_DSW}(a) shows the corresponding normalized free-energy landscape, $F/(N_{\mathrm{ch}}k_{\mathrm B}T_{\mathrm f})$, at the final temperature $T_{\mathrm f}=321~\mathrm{K}$ as a black solid line. The free energy exhibits a minimum at the collapsed equilibrium radius. Therefore, a particle initially located at the swollen radius (indicated by the blue square) relaxes downhill along the free-energy landscape towards the new equilibrium state (blue triangle). The slope of the free-energy curve determines the thermodynamic driving force acting on the swelling coordinate and therefore controls the rate at which the deswelling process takes place. As the microgel approaches its new equilibrium state, the absolute value of this slope progressively decreases. Consequently, the energetic driving force becomes weaker and the deswelling kinetics gradually slows down until the minimum of the free-energy landscape is reached.

Fig.~\ref{fig:F_DSW}(a) also illustrates the opposite temperature-jump protocol. In this case, the microgel is initially at equilibrium in the collapsed state, with $R(t=0)=200~\mathrm{nm}$ at
$T_{\mathrm{ini}}=321~\mathrm{K}$, and the temperature is suddenly decreased to $T_{\mathrm f}=288~\mathrm{K}$. The corresponding free-energy landscape at the final temperature is shown as a red dashed line. Under these conditions, the minimum of the free energy shifts towards larger radii, corresponding to the swollen equilibrium state with \(R_{\mathrm{eq}}=345~\mathrm{nm}\). The microgel therefore evolves towards increasing radius, driven by the thermodynamic force associated with the new free-energy landscape, until the swollen equilibrium state is reached.

It is also instructive to analyze the dependence of $D_{\mathrm{SW}}$ on the microgel radius. The black solid line in Fig.~\ref{fig:F_DSW}(b) shows $D_{\mathrm{SW}}(R,T_\mathrm{f})$ for the deswelling process. First, we note that the numerical values of $D_{\mathrm{SW}}$ are of the order of $10^{-17}~\mathrm{m}^2\,\mathrm{s}^{-1}$, i.e., about six orders of magnitude smaller than typical values of the collective diffusion coefficient $D_{\mathrm{coll}}$. This large difference arises because both quantities describe different dynamical processes. $D_{\mathrm{coll}}$ is a continuum transport coefficient that characterizes the propagation of local concentration or deformation perturbations through the gel network. By contrast, $D_{\mathrm{SW}}$ is an effective diffusion coefficient associated with the global swelling coordinate $R$, and it describes the mobility of the overall particle size along the free-energy landscape.

A particularly interesting feature is that $D_{\mathrm{SW}}$ is not monotonic with $R$. Instead, it reaches a maximum at an intermediate particle size, $R_{\mathrm{max}}=262.94~\mathrm{nm}$, corresponding to a polymer volume
fraction $\phi_{\mathrm{max}}=0.22$. This maximum arises from the non-monotonic dependence of the factor $\phi/\eta_{\mathrm{eff}}(\phi,T_{\mathrm f})$. Indeed, using Eq.~\eqref{eq:DSW}, one has $D_{\mathrm{SW}}\propto \phi/\eta_{\mathrm{eff}}(\phi,T_{\mathrm f})$ for fixed temperature. On the one hand, increasing $\phi$ corresponds to a more collapsed microgel and therefore to a denser polymer environment, which increases the effective viscosity experienced by the water molecules and tends to reduce $D_{\mathrm{SW}}$. On the other hand, the prefactor $\phi$ increases as the microgel radius decreases, reflecting the fact that the osmotic pressure increases because the thermodynamic force is distributed on a smaller surface. The competition between these two effects gives rise to the observed maximum.

The behavior is very similar for the reverse swelling process, represented by the red dashed line in Fig.~\ref{fig:F_DSW}(b). In this case, as the microgel radius increases, $D_{\mathrm{SW}}$ also crosses a maximum located at the same value of $\phi$, since the position of the maximum is controlled by the $\phi$-dependent part of the mobility. However, for a given value of $R$, the values of $D_{\mathrm{SW}}$ are larger for the deswelling process. This difference is due to the different final temperatures involved in the two temperature jumps. Since
$D_{\mathrm{SW}}\propto T_{\mathrm f}/\eta(T_{\mathrm f})$, the higher final temperature in the deswelling process leads to a larger value of $D_{\mathrm{SW}}$: the factor $T_{\mathrm f}$ is larger and the viscosity of bulk water is lower. Quantitatively, the ratio between the two curves is approximately
$[321/\eta(321)]/[288/\eta(288)]\simeq 2.25$.

\begin{figure}[ht!]
	\centering
\includegraphics[width=1.0\linewidth]{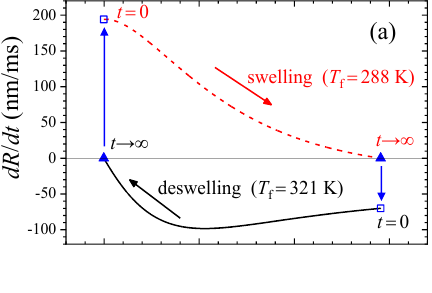}
	\caption{Instantaneous rate of change of the microgel radius, $dR/dt$, as a function of $R$ for the deswelling process (black solid line) and the swelling process (red dashed line).}
    \label{fig:dRdtnew}
\end{figure}

It is also instructive to analyze the instantaneous rate of change of the microgel radius, $dR/dt$, as a function of $R$ after the sudden temperature jump. This quantity is shown in Fig.~\ref{fig:dRdtnew} for both the swelling and deswelling processes. As expected, $dR/dt$ is positive during swelling, where the particle radius increases with time, and negative during deswelling, where the particle radius decreases. In both cases, the initial value of $dR/dt$ is finite, reflecting the abrupt onset of the size change immediately after the temperature jump. The rate then tends to zero as the microgel
approaches its final equilibrium state.

For the deswelling process, the absolute value of $dR/dt$ first increases progressively, indicating an initial acceleration of the collapse dynamics. The maximum deswelling rate is reached at approximately $R\simeq 253~\mathrm{nm}$. Beyond this point, the magnitude of $dR/dt$ decreases continuously, and the radius relaxation slows down until $dR/dt=0$ at the final collapsed equilibrium state. In the swelling process, a small maximum of $dR/dt$ is also observed very close to the initial collapsed state. After this initial stage, the swelling rate decreases monotonically as the microgel approaches the final swollen equilibrium radius.

The corresponding second time derivative (not shown), $d^2R/dt^2$, also reflects the nonlinear character of the relaxation. In both deswelling and swelling, it changes sign along the pathway and reaches an extremum at an intermediate stage, where the rate of change of $dR/dt$ is largest. Both $dR/dt$ and $d^2R/dt^2$ vanish at the final equilibrium state, confirming the overdamped character of the dynamics and the absence of oscillations around the final radius.

\begin{figure}[ht!]
	\centering
\includegraphics[width=1.0\linewidth]{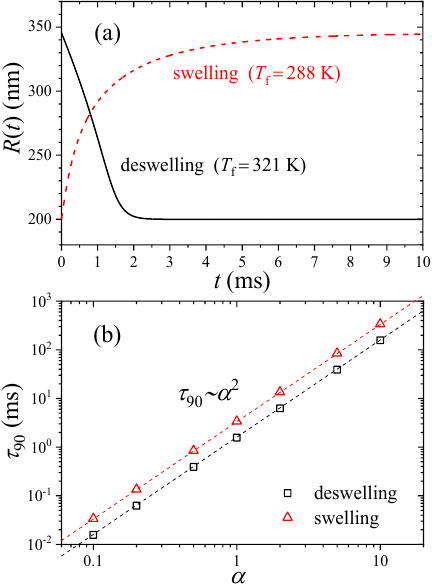}
	\caption{(a) Time evolution of the microgel radius, $R(t)$, during the deswelling and swelling processes. (b) Scaling of the characteristic relaxation time $\tau_{90}$ with the particle-size scaling factor $\alpha$ for both deswelling and swelling. The quantity $\tau_{90}$ is defined as the time required for the microgel radius to complete $90$\% of the total change between the initial and final equilibrium states.}
    \label{fig:Rt}
\end{figure}

The time evolution of the microgel radius as a function of time can be obtained from direct numerical integration of the kinetic differential equation:
\begin{equation}
    t=\int_{R(t=0)}^{R(t)}\frac{1}{M(R,T_\mathrm{f})}dR.
\end{equation}

The resulting time evolution of the microgel radius, $R(t)$, during the deswelling and swelling processes after the temperature jump is shown in Fig.~\ref{fig:Rt}(a). In both cases, the radius relaxes towards the corresponding final equilibrium value, although the shape of the relaxation curve is markedly different in the two processes. For deswelling, $R(t)$ exhibits an almost linear decrease over most of the trajectory, until the particle approaches the collapsed state. This behavior can be understood from the fact that the relaxation rate, $dR/dt$, remains nearly constant over a broad range of radii, giving rise to an approximately uniform decrease of the particle size. Only close to the collapsed equilibrium state does the relaxation become smoother, as the thermodynamic driving force progressively vanishes and $dR/dt$ tends to zero. The collapsed state is reached after approximately $2~\mathrm{ms}$.

By contrast, the swelling process displays a qualitatively different relaxation profile. In this case, the approach to the swollen equilibrium state is much more gradual and resembles a conventional exponential relaxation. The radius increases rapidly at early times and then slows down progressively as the microgel approaches the final swollen state. These results clearly show that swelling and deswelling are not kinetically symmetric processes. Instead, the microgel follows different relaxation pathways depending on the direction of the temperature jump, reflecting the state dependence of both the thermodynamic driving force and the swelling mobility along the relaxation trajectory.

To quantify this kinetic asymmetry, we define a characteristic swelling/deswelling time, $\tau_{90}$, as the time required for the microgel radius to complete $90$\% of the total change between the initial and final equilibrium states. Using this criterion, we obtain $\tau_{90}=1.609~\mathrm{ms}$ for the deswelling process and
$\tau_{90}=3.382~\mathrm{ms}$ for the swelling process. The relaxation times predicted by our model lie within the millisecond range reported for submicrometric thermoresponsive microgels, although the precise value depends strongly on the heating protocol, network architecture, and definition of the relaxation time. Indeed, time-resolved pressure-jump SANS measurements on acrylamide-based microgels with swollen radii of about $185~\mathrm{nm}$ have revealed volume-phase-transition kinetics with characteristic times of the order of milliseconds.~\cite{Wrede2018} Comparable time scales have also been reported by temperature-jump spectroscopy of Au--pNIPAM core--shell microgels, with values depending on shell thickness and cross-linker density.~\cite{Tagdell2022} In addition, our results confirm that swelling and deswelling are not kinetically symmetric processes. This prediction is qualitatively consistent with recent time-resolved experiments on pNIPAM-based nanogels, where collapse was found to occur much faster than reswelling after a temperature perturbation.~\cite{Wang2001,Murphy2016,Tagdell2022,Dallari2024}

One of the most important experimental signatures of microgel swelling kinetics is the quadratic dependence of the characteristic relaxation time on the particle size. To test whether our model reproduces this behavior, we performed numerical calculations of $\tau_{90}$ for geometrically scaled microgels. More specifically, the microgel radius was scaled by a factor $\alpha$, so that $R'=\alpha R$, while preserving the internal morphology of the polymer network. This means that the local polymer volume fraction and the reduced structural parameters of the network were kept fixed, namely $\phi$, $\phi_0$, $n$, and $n_\mathrm{e}$. Under this scaling, the collapsed volume and the number of chains scale as $V_0'=\alpha^3 V_0$ and $N_{\mathrm{ch}}'=\alpha^3 N_{\mathrm{ch}}$.

Figure~\ref{fig:Rt}(b) shows the resulting values of $\tau_{90}$ as a function of the scaling factor $\alpha$ for both the deswelling and swelling processes. In both cases, the characteristic relaxation time follows the scaling $\tau_{90}\sim \alpha^2$, in agreement with the experimentally observed quadratic dependence of the relaxation time on particle size. This scaling can be rationalized directly from Eq.~\eqref{eq:dRdtM}. For two geometrically similar microgels, the kinetic function satisfies $M(R')=M(\alpha R)=(1/\alpha)M(R)$. Therefore the time evolution of the scaled radius obeys \begin{equation}
\frac{dR'}{dt} = \alpha\frac{dR}{dt} = \alpha M(R) = \alpha M(R'/\alpha) = \alpha^2 M(R').
\end{equation}

Introducing the rescaled time $t^\prime=\alpha^2 t$ one obtains that the scaled microgel follows the same kinetic equation as the original one, namely $dR'/dt'=M(R',T_\mathrm{f})$. Consequently, the model naturally predicts that the characteristic swelling/deswelling time scales as $\tau'=\alpha^2\tau$, consistent with the Tanaka--Fillmore scaling and with theoretical, experimental, and simulation studies of gel and microgel swelling kinetics.~\cite{Tanaka1979,Suarez2006,wahrmund2009,Nikolov2018}

\subsection{Nonequilibrium swelling kinetics under finite thermalization times}

\begin{figure}[ht!]
	\centering
\includegraphics[width=1.0\linewidth]{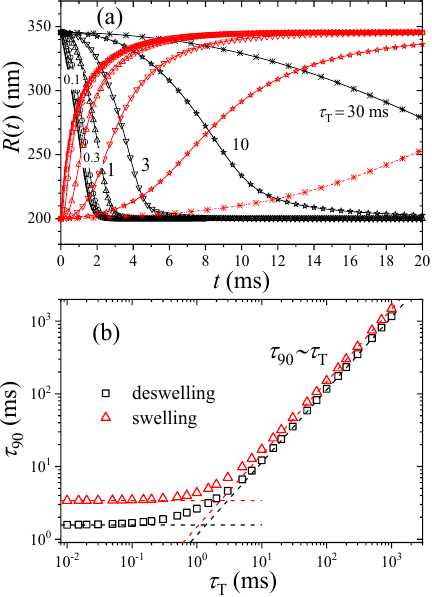}
	\caption{(a) Time evolution of the microgel radius, $R(t)$, during the deswelling and swelling processes. (b) Characteristic relaxation time $\tau_{90}$ as a function of $\tau_{\mathrm T}$ for both deswelling and swelling processes.}
    \label{fig:Rtprog}
\end{figure}

In the previous section, we analyzed the response of the microgel to an instantaneous temperature jump. However, in many experimental situations the thermalization of the surrounding medium is not instantaneous, and the temperature reaches its final value over a finite time scale. In this section, we consider this more general situation in order to investigate the competition between the intrinsic swelling/deswelling time of the microgel and the characteristic thermalization time of the external temperature field.

We assume that the temperature evolves exponentially from its initial value $T_{\mathrm{ini}}$ to its final equilibrium value $T_{\mathrm f}$ according to
\begin{equation}
    T(t)=T_{\mathrm f} +\left(T_{\mathrm{ini}}-T_{\mathrm f}\right)
    \exp\left(-t/\tau_{\mathrm T}\right),
\end{equation}
where $\tau_{\mathrm T}$ is the characteristic thermalization time. %For the deswelling process we take $T_{\mathrm{ini}}=288~\mathrm{K}$ and $T_{\mathrm f}=321~\mathrm{K}$, whereas for the swelling process we take $T_{\mathrm{ini}}=321~\mathrm{K}$ and $T_{\mathrm f}=288~\mathrm{K}$.
In the limit $\tau_{\mathrm T}\rightarrow 0$, this protocol reduces to the sudden temperature jump analyzed in the previous section. The corresponding values obtained for the instantaneous jump, $\tau_{90,\mathrm{jump}}=1.609~\mathrm{ms}$ for deswelling and $\tau_{90,\mathrm{jump}}=3.382~\mathrm{ms}$ for swelling, provide natural reference time scales. We therefore explore values of $\tau_{\mathrm T}$ both below and above these characteristic times in order to cover the different kinetic regimes.

The microgel radius as a function of time is obtained integrating numerically the differential equation $dR/dt=M(R,T(t))$. Lines with symbols in Fig.~\ref{fig:Rtprog}(a) show $R(t)$ for $\tau_{\mathrm T}=0.1$, $0.3$, $1$, $3$, $10$, and $30~\mathrm{ms}$. Black curves correspond to deswelling, whereas red curves correspond to swelling. The thicker solid lines show the limiting case of an instantaneous temperature jump. For all values of $\tau_{\mathrm T}$, swelling is slower than deswelling. When $\tau_{\mathrm T}\ll\tau_{90,\mathrm{jump}}$, thermalization is much faster than the intrinsic size relaxation of the microgel, as illustrated by the curves for $\tau_{\mathrm T}=0.1$ and
$0.3~\mathrm{ms}$. In this regime, the radius relaxation is very close to that obtained for the sudden temperature jump. As $\tau_{\mathrm T}$ increases, the kinetics becomes progressively slower and the shape of the relaxation curves is modified. In the opposite limit, $\tau_{\mathrm T}\gg\tau_{90,\mathrm{jump}}$, for example for $\tau_{\mathrm T}=30~\mathrm{ms}$, the radius evolves much more slowly. Moreover, the deswelling and swelling curves become more similar in shape, suggesting that the size change is then mainly dictated by the imposed temperature evolution rather than by the intrinsic relaxation dynamics of the microgel.

Fig.~\ref{fig:Rtprog}(b) confirms this crossover. For $\tau_T\ll\tau_{90,\mathrm{jump}}$, the relaxation time reaches a plateau corresponding to the intrinsic swelling or deswelling time after an instantaneous temperature jump. As $\tau_T$ becomes comparable to this intrinsic time, $\tau_{90}$ departs from the plateau. In the opposite limit, $\tau_T\gg\tau_{90,\mathrm{jump}}$, the relaxation becomes thermalization-controlled and $\tau_{90}$ increases linearly with $\tau_T$, indicating that the microgel follows the slowly varying equilibrium state almost quasi-statically.
\begin{figure}[ht!]
	\centering
\includegraphics[width=1.0\linewidth]{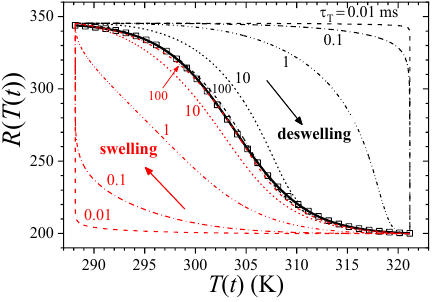}
	\caption{Nonequilibrium microgel radius, $R$, as a function of the instantaneous temperature, $T$, for different values of the thermalization time $\tau_{\mathrm T}$. Black curves correspond to the deswelling protocol, whereas red curves correspond to the swelling protocol. The square symbols represent the equilibrium swelling curve, $R_{\mathrm{eq}}(T)$. The separation between the two branches reflects a dynamic nonequilibrium delay caused by the finite response time of the microgel.}
    \label{fig:histeresis}
\end{figure}

An even more direct way to visualize the departure from equilibrium is to eliminate the time variable and represent the instantaneous microgel radius as a function of the instantaneous temperature, $R=R(T)$. The resulting curves are shown in Fig.~\ref{fig:histeresis} for different values of the thermalization time $\tau_{\mathrm T}$, both for the deswelling and swelling protocols.

For large values of $\tau_{\mathrm T}$, the temperature changes slowly compared with the intrinsic swelling/deswelling dynamics of the microgel. As a result, the particle remains close to equilibrium throughout the process, and the nonequilibrium curves approach the equilibrium swelling curve $R_{\mathrm{eq}}(T)$, represented by square symbols. As $\tau_{\mathrm T}$ decreases, the microgel progressively deviates from this equilibrium curve, revealing an increasing delay of the size response with respect to the imposed temperature change. In the limit of very short thermalization times, the temperature reaches almost its final value before the microgel has had time to respond. For the deswelling protocol, this produces an almost horizontal trajectory at the initial swollen radius during most of the temperature increase, followed by a rapid collapse only once the temperature is already close to $T_{\mathrm f}=321~\mathrm{K}$. A similar delayed response is observed for the swelling protocol after the temperature is decreased.

The deswelling and swelling trajectories therefore form a dynamic nonequilibrium loop. In contrast to equilibrium or morphological hysteresis in volume phase transitions of gels,~\cite{Sekimoto1993} this loop originates from the finite response time of the microgel combined with the finite thermalization time of the external temperature field. Its area decreases in the quasi-static limit and increases when the imposed temperature variation is fast compared with the intrinsic microgel relaxation. This result has a direct implication for experiments: if thermalization is not accounted for, the measured radius relaxation may be incorrectly attributed entirely to intrinsic microgel dynamics, leading to an overestimation of the swelling or deswelling time. In particular, when $\tau_{\mathrm T}$ is comparable to or larger than the intrinsic swelling/deswelling time, the observed relaxation is strongly affected, or even dominated, by the externally imposed temperature evolution. This conclusion is consistent with recent ultrafast measurements of PNIPAM-based core--shell nanogels, where the extracted collapse and swelling times were shown to depend on the interplay between local heating, cooling, and the intrinsic response of the polymer shell.~\cite{Dallari2024} Related temperature-jump experiments based on localized nanoscale heating have also reported ultrafast particle collapse, several orders of magnitude faster than the response obtained under conventional heating.~\cite{Zhao2019}

\section{Smoluchowski description of nonequilibrium size fluctuations}
\label{sec:theory_Smoluchowski}

Finally, in this section we extend the deterministic model discussed previously to a stochastic description of the swelling coordinate, allowing us to include thermal fluctuations of the microgel radius. For this purpose, we introduce  the probability density $P(R,t)$, which satisfies the normalization condition $\int P(R,t)dR=1$, and obeys the Smoluchowski partial differential equation obtained as a stochastic generalization of Eq.~\eqref{eq:RtDSW}:~\cite{Dhont1996}
\begin{equation}
\label{eq:Smoluchowski}
\frac{\partial P}{\partial t}=\frac{\partial}{\partial R}
\left[D_{\mathrm{SW}}(R)\left(
\frac{\partial P}{\partial R}+
P\frac{\partial \beta F}{\partial R}
\right)
\right].
\end{equation}

For fixed external conditions, Eq.~\eqref{eq:Smoluchowski} approaches the equilibrium distribution $P_{\mathrm{eq}}(R) \propto \exp\left[-\beta F(R)\right]$, which represents an intrinsic polydispersity of the microgel exclusively caused by the fluctuations of $R$. This stochastic extension therefore allows us to predict not only the time evolution of the average microgel radius, $\langle R(t) \rangle$, but also the complete particle-size distribution, its variance, and its transient evolution following a change in the external conditions. This stochastic description reduces to the deterministic equation given by Eq.~\eqref{eq:RtDSW} when thermal fluctuations are neglected, or equivalently in the limit $k_{\mathrm B}T\to0$ at fixed $\zeta_{\mathrm{SW}}=k_\mathrm{B}T/D_{\mathrm{SW}}$, defined as the effective swelling friction coefficient associated with changes in the global radius of the polymer network.
\begin{figure}[ht!]
	\centering
\includegraphics[width=1.0\linewidth]{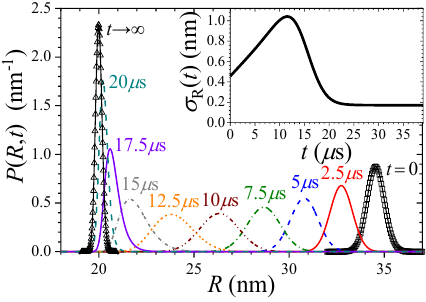}
	\caption{Time evolution of the probability density distribution $P(R,t)$ during the deswelling process obtained from the numerical solution of the Smoluchowski equation. The initial and final equilibrium distributions are shown with square and triangular symbols, respectively. Inset: time evolution of the standard deviation of the size distribution, $\sigma_R(t)$, revealing a non-monotonic behavior with a transient maximum during the collapse process.}
    \label{fig:rho_t}
\end{figure}

In order to make the probability distributions easier to visualize, we consider a microgel whose size is reduced by a factor of ten, corresponding to a scaling factor $\alpha=0.1$, while preserving the internal morphology of the polymer network. Under this scaling, the total number of polymer chains decreases by a factor $\alpha^3=10^{-3}$. As a result, the free-energy landscape becomes less steep and the equilibrium probability distribution becomes broader, making the thermal fluctuations of the radius more clearly visible. At the same time, the collapse kinetics is accelerated by a factor $\alpha^2=10^{-2}$, as demonstrated above. The Smoluchowski partial differential equation was solved numerically using a time step $\Delta t=10^{-8}~\mathrm{ms}$ and a spatial grid spacing $\Delta R=0.02~\mathrm{nm}$, imposing zero probability flux at $R=17~\mathrm{nm}$ and $R=40~\mathrm{nm}$. In the following, we focus only on the deswelling process after a sudden temperature jump from $288~\mathrm{K}$ to $T_\mathrm{f}=321~\mathrm{K}$.

Fig.~\ref{fig:rho_t} shows the time evolution of $P(R,t)$ during the deswelling process. The initial and final equilibrium distributions are shown as lines with square and triangular symbols, respectively. They are given by $P(R,0) = C_{\mathrm{ini}} \exp\left[-\beta_{\mathrm{ini}}F(R,T_{\mathrm{ini}})\right]$ and
$P(R,\infty)=C_{\mathrm f}
\exp\left[-\beta_{\mathrm f}F(R,T_{\mathrm f})\right]$, where $\beta_{\mathrm{ini}}=1/(k_{\mathrm B}T_{\mathrm{ini}})$ and
$\beta_{\mathrm f}=1/(k_{\mathrm B}T_{\mathrm f})$. The constants $C_{\mathrm{ini}}$ and $C_{\mathrm f}$ are normalization factors chosen so that the corresponding probability distributions integrate to unity. As observed, after the temperature jump the maximum of the probability distribution progressively shifts towards smaller radii, until the final equilibrium distribution centered around the collapsed radius is reached at long times.

A particularly interesting feature is that the height and width of the distribution evolve non-monotonically during the relaxation. At early times, as the microgel begins to collapse, the distribution broadens and its maximum height decreases. At later times, this trend is reversed: the distribution becomes progressively narrower and more peaked as it approaches the final collapsed equilibrium state. This behavior is quantified in the inset of Fig.~\ref{fig:rho_t}, where we plot the standard deviation of the size distribution, $\sigma_R(t)=(\langle R^2(t)\rangle-\langle R(t)\rangle^2)^{1/2}$, as a function of time. Initially, the distribution has a width $\sigma_R(0)=0.456~\mathrm{nm}$. During the first stage of the collapse, $\sigma_R(t)$ increases and reaches a maximum value $\sigma_{R,\mathrm{max}}=1.041~\mathrm{nm}$ at $t_{\mathrm{max}}=11.5~\mu\mathrm{s}$, when the mean radius is $\langle R\rangle\simeq 24.9~\mathrm{nm}$. After this point, the width decreases and eventually reaches the final equilibrium value $\sigma_R(\infty)=0.172~\mathrm{nm}$.

The transient maximum of $\sigma_R(t)$ results from the competition between enhanced diffusive spreading and deterministic confinement by the free-energy landscape. During the first stage of collapse, the probability distribution moves through an intermediate range of radii where $D_{\mathrm{SW}}(R)$ is large. In this region, fluctuations of the swelling coordinate are amplified and the distribution broadens in radius space. As the microgel approaches the collapsed state, however, the free-energy minimum becomes increasingly confining and the deterministic drift compresses the distribution towards the final equilibrium radius. This causes $\sigma_R(t)$ to decrease at later times. Therefore, the maximum of $\sigma_R(t)$ does not necessarily coincide with the maximum of $D_{\mathrm{SW}}(R)$: the distribution width is controlled by both local diffusive spreading and the drift imposed by the evolving free-energy landscape. The slight shift between the two maxima reflects the fact that the distribution continues to broaden until diffusive spreading is balanced by the confining drift towards the collapsed equilibrium state.

It is worth noting that the maximum of $\sigma_R(t)$ occurs when the mean radius lies within the transition region between the swollen and collapsed states. This indicates that the largest fluctuations of the global microgel size are generated while the particle crosses the intermediate configurations associated with the volume phase transition. In this region, the microgel is particularly susceptible to small thermodynamic perturbations, since relatively small changes in temperature or polymer volume fraction produce large changes in particle size. This prediction is qualitatively consistent with experimental observations on thermoresponsive pNIPAM-based microgels.~\cite{Tanaka1985,Tanaka1986} Scattering and microscopy studies have shown that the volume phase transition is accompanied by pronounced internal structural rearrangements, involving a transformation from a swollen, fuzzy, solvent-rich particle to a more compact collapsed state.~\cite{Kratz2001,Fernandez-Barbero2002} Moreover, mechanical measurements have shown that pNIPAM microgels become softer and more compressible in the vicinity of the volume phase transition.~\cite{Hansing2018,Sierra-Martin2011} 

In our model, the transient maximum of $\sigma_R(t)$ is a genuine nonequilibrium effect arising from the non-monotonic behavior of $D_{\mathrm{SW}}(R)$. When the distribution crosses this region of enhanced mobility, it broadens transiently before being compressed again by the final free-energy minimum, a prediction that could be tested experimentally using time-resolved DLS or single-particle tracking following a rapid temperature jump.

\section{Conclusions}
\label{sec:conclusions}

In this work, we have developed a solvent-flux theory for the nonequilibrium swelling and deswelling kinetics of thermoresponsive microgels. The central idea of the model is to describe the time evolution of the microgel radius in terms of the net solvent transport across the particle boundary, driven by the osmotic-pressure imbalance between the instantaneous microgel state and the surrounding solvent. By combining this flux-based description with a Flory--Rehner free-energy model and a state-dependent water mobility inside the polymer network, we obtained a closed nonlinear dynamical equation for the microgel radius.

A central outcome of the theory is the derivation of an explicit state- and temperature-dependent swelling mobility, $D_{\mathrm{SW}}(\phi,T)$. This quantity governs the dynamics of the global swelling coordinate and should not be identified either with the molecular self-diffusion coefficient of water or with the collective diffusion coefficient appearing in the Tanaka--Fillmore description. In the linear-response limit, the present model reduces to an exponential relaxation and recovers the characteristic Tanaka--Fillmore size scaling, $\tau_{\mathrm{SW}}\sim R_{\mathrm{eq}}^2\gamma/K_{\mathrm{eq}}$. Within this limit, the theory also provides a microscopic interpretation of the phenomenological polymer--solvent friction coefficient, $\gamma$, in terms of solvent transport through the polymer network. Beyond linear response, the state dependence of $D_{\mathrm{SW}}$ leads to asymmetric swelling and deswelling trajectories, with deswelling being faster than swelling. The model also reproduces the quadratic size scaling observed experimentally in microgel swelling kinetics.

We have also shown that finite thermalization times can strongly affect the apparent nonequilibrium swelling and deswelling kinetics. When the external temperature changes much faster than the intrinsic microgel relaxation time, the response approaches the sudden-jump limit. However, when the thermalization time becomes comparable to or larger than the intrinsic swelling/deswelling time, the observed radius evolution is increasingly controlled by the imposed temperature protocol. In this regime, the microgel follows the slowly varying equilibrium state almost quasi-statically, and the radius--temperature trajectories form dynamic hysteresis-like loops. This result highlights the importance of accounting for thermalization effects when extracting intrinsic microgel relaxation times from experiments.

Finally, we extended the deterministic theory to a stochastic Smoluchowski description that incorporates thermal fluctuations of the microgel radius. For the deswelling process, the theory predicts a non-monotonic evolution of the distribution width: the size distribution first broadens while the microgel crosses the volume-transition region and then narrows as the final collapsed state is approached. This transient maximum of the fluctuations is a nonequilibrium effect associated with the state-dependent and non-monotonic behavior of $D_\mathrm{SW}(R)$, together with the subsequent confinement by the final free-energy minimum.

Overall, the present theory provides a physically transparent framework for connecting solvent transport, free-energy landscapes, nonlinear nonequilibrium kinetics, finite thermalization effects, and thermal size fluctuations in responsive microgels. Beyond the specific pNIPAM system considered here, the approach could be extended to more complex microgel architectures, including partially inhomogeneous networks (core--shell morphology),~\cite{moncho-jorda2025} charged microgels under varying ionic strength, and collective swelling effects in concentrated interacting suspensions.~\cite{Holmqvist2012}

\acknowledgments

A.M.-J and A.P. acknowledge funding grant No. PID2022-136540NB-I00 awarded by MICIU/AEI/10.13039/501100011033 and ERDF, a way of making Europe. A.P. also acknowledges Grant W911NF-23-1-0099 awarded by the U.S. Army Research Office. We thank PROTEUS, the supercomputing center of the Institute Carlos I of Theoretical and Computational Physics of the University of Granada and C3UPO of the Universidad Pablo de Olavide for the support with HPC facilities.

% If in two-column mode, this environment will change to single-column format so that long equations can be displayed. 
%\begin{widetext}
%$$\mbox{put long equation here}$$
%\end{widetext}

% Figures should be put into the text as floats. 
% Use the graphics or graphicx packages (distributed with LaTeX2e).
% Put the label that you will use with \ref{} command in the braces of the \label{} command.
%
% \begin{figure}
% \end{figure}

% \begin{table}
% \end{tabular}
% \end{table}

%\begin{acknowledgments}
% Put your acknowledgments here.
%\end{acknowledgments}

\bibliography{swelling_bib}

\end{document}